\documentclass{llncs}
\usepackage{frameit,amsmath,color,latexsym,graphics,wrapfig,amssymb,balance}

\usepackage{times,epsfig}
\usepackage{graphicx}

\usepackage{ifthen}
\usepackage{times,epsfig,amsmath,clrscode3e}
\usepackage{frameit,amsmath,color,latexsym,graphics,wrapfig,textcomp,dsfont,centernot,enumitem,mathtools}
\usepackage[utf8]{inputenc}

\usepackage{hyperref}
\usepackage{url}
\usepackage {tikz}
\usetikzlibrary{arrows,automata}

\usepackage[linesnumbered,ruled,vlined]{algorithm2e}

\newcommand{\setvars}[1]{\ensuremath{\bar{#1}}}

\newcommand{\savespace}{\vspace{-2mm}}
\newcommand{\saveone}{\vspace{-1mm}}

\mathchardef\mhyphen="2D

\newcommand{\seppredF}[2]{\ensuremath{#1(#2)}}
\newcommand{\seppred}[2]{\ensuremath{#1(#2)}}

\newcommand{\base}{\ensuremath{{\pure^b}}}
\newcommand{\ind}{\ensuremath{{\pure}}}

\newcommand{\report}[1]{ }

\newcommand{\acm}[1]{ }

\newcommand{\hide}[1]{}
\newcommand{\hideie}[1]{}

\newcommand{\pure}{\ensuremath{\pi}}
\newcommand{\seq}{\ensuremath{E}}
\newcommand{\astart}{\ensuremath{\Upsilon}}
\newcommand{\subterm}{\ensuremath{\Lambda}}

\newcommand{\myit}[1]{\textit{#1}}

\newcommand{\unf}[1]{[\underline{{\bf \scriptstyle UNF-#1}}]}

\def\FV{\myit{FV}}

\def\D{\Delta}

\def\max{\myit{max}}
\def\min{\myit{min}}
\def\fresh{\myit{fresh}}

\def\int{\code{int}}

\def\true{\code{true}\,}
\def\false{\code{false}\,}
\def\a{a}

 \newcommand{\code}[1]{{\small {\ensuremath{\tt #1}}}}

\newcommand{\btt}[1]{{\ensuremath{\tt #1}}}

\newcommand{\locnay}[1]{}
\newcommand{\wnnay}[1]{}
\newcommand{\longnay}[1]{}
\newcommand{\nodo}[1]{}

\newcommand{\defpred}{\btt{pred}}

\newtheorem{thm}{Theorem}[section]
\numberwithin{thm}{section} 

\newtheorem{defn}{Definition}

\def\S{\Sigma}

\def\SStore{\myit{SStacks}}
\def\IStore{\myit{ZStacks}}

\def\iffs{\small \btt{ iff~}}

\def\fresh{\myit{fresh}}

\newcommand{\atom}{\alpha}

\newcommand{\yields}{\leadsto}

\newcommand{\defsym}{\ensuremath{\overset{\text{\scriptsize{def}}}{=}}}

\newcommand{\sstack}{\ensuremath{\eta}}
\newcommand{\istack}{\ensuremath{\beta_{\eta}}}
\newcommand{\force}{\ensuremath{\models}}

\newcommand{\form}[1]{\ensuremath{#1}}

\newcommand{\regex}{\mathcal{R}}

\newcommand{\basetheory}{\ensuremath{\mathcal{L}}}

 \newcommand{\sat}{{\sc{\code{SAT}}}}
 \newcommand{\unsat}{{\sc{\code{UNSAT}}}}

\def\qed{\hfill\ensuremath{\square}}

\newcommand{\subst}[2]{\ensuremath{[#1 {/} #2]}}

\newcommand{\satb}[1]{\code{\bf sat^b}(\ensuremath{#1})}

\newcommand{\utree}[1]{\ensuremath{{\cal T}_{#1}}}

\newcommand{\unfoldsolve}{\code{unfold_{\solvedF}}}

\newcommand{\uosolve}{\code{UA_{\solvedF}}}
\newcommand{\oosolve}{\code{OA_{\solvedF}}}

\newcommand{\alphabet}{\Sigma}
\newcommand{\semp}{\epsilon}
\def\SVar{\myit{U}}
\def\IVar{\myit{I}}
\newcommand{\lang}{\ensuremath{\mathcal{L}}}
\newcommand{\strel}{{\tt SEA}}

\def\se{\myit{E}}
\newcommand{\seqs}{\ensuremath{Es}}
\def\ses{\myit{Es}}

\newcommand{\solvedF}{{\ensuremath{\tt 1\strel}}}
\newcommand{\simpleF}{{\ensuremath{\tt {0\strel}}}}
\newcommand{\satsolved}{{\ensuremath{\tt S2_{1\strel}}}}

\newcommand{\satalgo}{{\sc{\code{{S2SAT}}}}}

\newcommand{\lbsolve}{{\code{fp_{1SEA}}}}

\newcommand{\conferencepaper}{1} 
\newcommand{\rep}[1]{\ifthenelse{\conferencepaper = 0}{#1}{}}
\newcommand{\repconf}[2]{\ifthenelse{\conferencepaper = 0}{#1}{#2}}
\begin{document}

\title{A Decision Procedure for String Logic with
 Equations, Regular Membership and Length Constraints}

\author{
Quang Loc Le
 }
 \institute{
 Singapore University of Technology and Design
 }


\maketitle

\begin{abstract}

In this paper, we consider the satisfiability problem for string logic
with equations, regular membership and Presburger constraints over length functions.
The difficulty comes from multiple occurrences of string variables
making state-of-the-art algorithms non-terminating.
Our main contribution is to show that the satisfiability problem in a fragment where
 no string variable occurs more than twice in an equation
is decidable.
In particular, we propose
 a semi-decision procedure for arbitrary string formulae with
word equations, regular membership and length functions.
The essence of our procedure is an
algorithm to
enumerate an equivalent set of {\em solvable} disjuncts for the formula.
We further show that the algorithm always terminates
for the aforementioned decidable fragment.
Finally, we
provide a complexity analysis of our decision procedure
to prove that it
runs, in the worst case, in factorial time.

\keywords String logic {$\cdot$} Satisfiability {$\cdot$} Decision Procedure
{$\cdot$} Inductive Predicates

\end{abstract}

\section{Introduction}\label{sec.intro}
There has been significant recent interest in
reasoning about web and database programs for bug finding \cite{Artzi:TSE:2010}
and vulnerability verification \cite{Kiezun:ISSTA:2009} due to
a huge number of security threats over the Internet.
In these reasoning systems, solvers
 for constraint languages over strings (a.k.a. string solvers)
 plays a central role.
The problem of solving word equations had been established.
In 1977, Makanin notably proved that 
the satisfiability problem of word equations is decidable  \cite{Makanin:math:1977}.
Following up the great Makanin's seminal paper, 
many studies either
improved complexity for this algorithm
\cite{Koscielski:JACM:1996,Plandowski:FOCS:1998,Plandowski:FOCS:1999} or search for a minmal and complete set of solutions \cite{Jaffar:JACM:1990,Plandowski:STOC:2006}.
However, reasoning about web applications and database programs
typically requires a constraint language
including word equations, regular membership
and arithmetic on length functions.
As an example, a function which generates new user accounts
is often required to validate validity of user-name
 (whether it contains some special characters i.e., '@')
and password (whether its length is longer than a certain number i.e., 8).
Since the length constraints implied by a word equation
is not always represented with finitely many equations in numeric form described by Plandowski
  \cite{Plandowski:STOC:2006},
developing a decision procedure for the combined theories
 is not straightforward.

There has been a few studies on
foundations for string formulas 
which combine
 word equations, regular membership and length constraints.
Ganesh {\em et. al.} 
presented decidability result
for the combination of
word equations and linear arithmetic \cite{Ganesh:HVF:2012}.
The formulas in this fragment 
are restricted such that no string variable occurs twice in an equation.
Abdulla {\em et. al.} further extended the result
with regular membership to acyclic fragment
 \cite{Abdulla:CAV:2014}.
Liang {\em et. al.} formalized the acyclic fragment
without word equations
using the calculus in \cite{Liang:FroCos:2015}.
Finally, Ganesh {\em et. al.} have recently shown
the undecidability of the satisfiability problem for
the
theories over string equations, length function,
 and string-number conversion predicate \cite{DBLP:journals/corr/GaneshB16}.
So far, there is no decision procedure
 supporting for a 
 fragment of word equations
 and length functions
beyond the acyclic fragment discussed above.

Practical approaches to solving constraints of
 string logic have been developed dramatically.
Initial approaches \cite{Hooimeijer:PLDI:2009,Hooimeijer:ASE:2010,Veanes:ICST:2010,Yu:SAS:2010}
 which are based on automata
  have difficulties in handling
string constraints related to length functions.
To overcome this problem,
 bounded approaches -
automata-based \cite{Kiezun:ISSTA:2009,Axelsson:ICALP:2008,He:CP:2013}
as well as bit vector-based \cite{Bjorner:TACAS:2009,Saxena:SP:2010} -
 support those queries whose
 string variables have bounded lengths.
 These approaches could efficiently support for satisfiability ({\sat}).
 However, they may not be sound for  unsatisfiability ({\unsat}).
Recently, unbounded approaches \cite{Zheng:FSE:2013,Liang:CAV:2014,Trinh:CCS:2014,Trinh:CAV:2016,Zheng:CAV:2015} support words as primitive type and are
successfully integrated into Satisfiability Modulo Theories framework.
The main technique used in these solvers is ``Unfold-and-Match''
which is to incrementally reduce the size of the input, via splitting
and/or unfolding process.
Although this technique is effective and efficient for a large number of
queries over the
 combined theories of string and arithmetic,
it does not work
for those
queries which have more than one occurrence
of every string variables.
For instance, the solvers \cite{Zheng:FSE:2013,Abdulla:CAV:2014,Liang:CAV:2014,Trinh:CCS:2014,Abdulla:CAV:2015}
 did not terminate when deciding satisfiability for
 the following formula which has two occurrences of the string variable \form{s}:
\[
\form{\pure~{\equiv}~a~{\cdot}~b~{\cdot}~s{=}s~{\cdot}~b~{\cdot}a~}
\]
For efficiency, new heuristics has recently introduced in \cite{Zheng:CAV:2015}
and \cite{Trinh:CAV:2016} to avoid such non-termination.
However, these approaches are not complete.
Our main contribution is a decision procedure for the constraint
language including the formula \form{\pure} above.

In this work,
we present a new semi-decision procedure, called {\satsolved},
for
a fragment of string logic, called {\strel}, which
includes
 word
equations, regular membership and arithmetical constraints over
 length functions.
The proposed procedure provides
an answer, which is either {\sat} (with a model, a valuation assignment to variables of the input) or {\unsat}, for the satisfiability problem.
Different to the existing approaches, we propose  {\em inductive predicate} to model
 string variable together with length function.
The core idea of {\satsolved} is an
algorithm to enumerate the complete set of solutions 
 for a given {\strel} formula. 
Each solution is {\em solvable}
i.e., is defined in a sound and complete
{\em base} logic, called {\simpleF} fragment.

{\satsolved} takes  a  formula in {\strel} logic
 as input.
It iteratively constructs a series of {\em unfolding trees} for the input
by unfolding inductive predicates in a
complete manner
until either a {\sat} leaf or
a proof of {\unsat} is identified.
In each iteration, it
examines every leaves of the tree (the disjunction of which is equivalent to the input formula)
with under-approximation, over-approximation and back-link construction
for cyclic proofs.
 In particular, 
  {\satsolved} first checks satisfiability for leaves
which are in the base logic. 
 These leaves are under-approximation of the input and
are precisely decided.
 Second,
{\satsolved} over-approximates open (non-unsatisfiable) leaves
  prior to checking their unsatisfiability.
Next, remaining open leaves are either
linked back to an interior nodes (to form a partial cyclic proof).
Leaves which are either unsatisfiable, or linked are marked closed.
 Otherwise, they are
 open.
Finally, if all leaves are closed then  {\satsolved} returns {\unsat}.
Otherwise, it chooses
 an open leaf in a depth-first manner
for unfolding
 inductive predicates,
 matching and moving to the next iteration.
For unfolding, {\satsolved}
 applies an Unfold-and-Match strategy
  on the leading terms (either string variables
or constant characters) of the left-hand-side (LHS)
and right-hand-side (RHS) of a word equation.

Our main contribution
is a decidable subfragment,
called {\solvedF}, so as
 the proposed procedure always terminates.
 There are two restrictions on {\solvedF} formulas.
The first restriction is that
  either (i) no string variable occurs twice in an equation
  or (ii)
 no string variable occurs more than twice in an equation
 with some additional restrictions in arithmetic.
The second restriction applied on formulas with
 multiple word equations
is that every formulas deduced by {\satsolved} satisfy
 the first restriction.
Our Unfold-and-Match strategy ensures that notational length of the equation
 decreases at least one for type (i) formulas and
does not increase for type (ii) formulas.
This makes {\satsolved} solver terminating
for formulas in the  {\solvedF} fragment.
We undertake a complexity analysis of our decision procedure
which shows that, in the worst case, it runs in
 linear time for type (i)
and in factorial time for type (ii) of {\solvedF}.

\paragraph{Contributions.} We make the following primary contributions.
\begin{itemize}
\item We propose semi-decision procedure {\satsolved} for word equations,
regular expression and arithmetic constraints on length functions.
\item We present a subfragment where {\satsolved} always terminates
and thus becomes a decision procedure.
\item We provide computational complexity results for the satisfiability
 on the decidable fragments. 
\end{itemize}


\section{Preliminaries}
In this section, we present the string logic {\strel}.
We also describe a normalized form
which our solver is built upon.

\subsection{{\strel} String Logic}
Concrete string models assume a finite alphabet \form{\alphabet},
set of finite words over \form{\alphabet^*}, and
a set of integer numbers \form{Z}.
 We work with a set \form{\SVar}
of string variables denoting words in \form{\alphabet^*},
and a set \form{\IVar} of arithmetical variables.

 \begin{figure}[tb]
 \begin{center}
  \begin{minipage}{0.85\textwidth}
  \begin{frameit} \savespace 
\[
\begin{array}{c}
\begin{array}{lcll}
\text{disj formula} & \pure &::=& \phi ~|~ \phi_1 ~{\vee}~ \phi_2  \\
\text{formula} & \phi &::=&    \atom \mid \myit{i}
                \mid tr{\in} \regex
                 ~|~ \neg\phi_1 \mid
\phi_1 {\wedge} \phi_2  \\
\text{(dis)equality} & \atom &::=&  tr_1 {=} tr_2 ~|~ tr_1 {\ne} tr_2 \\
\text{term} & tr &::=&  \semp ~|~ c \mid s \mid \seppredF{\code{STR}}{u{,}n}
 \mid tr \cdot tr\\
\text{regex} & \regex &::=&  \emptyset \mid \semp ~|~ c \mid w
 \mid \regex \cdot \regex \mid \regex {+} \regex \mid 
   \regex \cap \regex \mid \regex^C \mid \regex^*\\
 \text{Presburger}
 &\myit{i} &::=& \a_1 {=} \a_2 \mid \a_1 {\leq} \a_2 \\
 & a &::=& \!\!
    \begin{array}[t]{l}
      k^{\int} |~ v \mid |s| \mid k^{\int}{\times}\a  \mid {-}\a |~
\a_1 {\%} \a_2 ~|\\
       \a_1 \!+\! \a_2 \mid \max(\a_1{,}\!\a_2) \mid \min(\a_1{,}\!\a_2)
    \end{array}\\
 \end{array} \\
\begin{array}{l}
  c \in {\alphabet} \quad w{,}y{,}z \in {\alphabet}^* \quad s{,}u{,}t \in {\SVar} \quad
 v{,}x \in \text{\IVar} \quad k^{int} \in \text{Z}
\end{array}
\end{array}
\]
 \end{frameit}
 \end{minipage}
\caption{Syntax}\label{prm.spec.fig}
\end{center} \savespace 
\end{figure}

\paragraph{Syntax}
The syntax of quantifier-free string formulas in {\strel}
 is presented in Fig. \ref{prm.spec.fig}.
Regular expressions \form{\regex} does not contains any string variables.
We use \form{\se} to denote a word equation
and \form{\seqs} a conjunctive sequence of word equations.
\form{\seqs_i} to denote the $i^{th}$ word equation in the sequence.
We use \form{w_{|k|}} for
an arbitrary word in \form{\alphabet^*} with length \form{k},
 and
 \form{w^n} to denote
the word which is a concatenation of \form{n} word \form{w},
i.e. \form{w^n{\equiv}w{\cdot}...{\cdot}w} (\form{n} copies).
We use \form{\pure\subst{t1}{t2}} for a
substitution of  all occurrences
of \form{t_2} in \form{\pure} to \form{t_1}.
%
We define inductive predicate \code{STR} to encode
string variables
as follows.
 \begin{defn}[\code{STR} Predicate] A string variable is defined via
 the inductive predicate \code{STR} as:
\form{ \seppred{\code{STR}}{u{,}n}  ~{\equiv}~ u{=}\semp {\wedge}n{=}0
~{\vee}~   \seppred{\code{STR}}{u_1{,}n_1}
 {\wedge}u{=}c{\cdot}u_1 {\wedge}
 n_1{=}n{-}1 {\wedge} n{>}0},
where
\form{u} and \form{n} are parameters:
 n is the length of string variable \form{u} and
\form{c{\in}\alphabet}. 
\end{defn}
This predicate has the invariant \form{n{\geq}0}.
In the inductive rule, \form{u_1} is a subterm of \form{u}
and \form{u{=}c{\cdot}u_1} is a subterm constraint.
This subterm is important for cyclic proof to detect isomorphic word equations.
A string variable may be in bare form (without a \code{STR} predicate) or
 \code{STR} predicate instance.
We emphasize that \code{STR} instances are generated
and used by our solver.
They do not appear in the user-provided formulas.
We inductively define {length function of a string term \form{tr},
denoted as \form{|tr|}, 
as follows.}
\[
|\semp|{=}0 ~~~~|c|{=}1 ~~~~|w_{|k|}|{=}k~~~~
|\seppred{\code{STR}}{u{,}n}|{=}n
~~~~
|tr_1{\cdot}tr_2|{=}|tr_1|{+}|tr_2|
\]

\hide{In the following, we define the function \code{unfold\_str}
used to derive an equivalent disjunction
of a formula through unfolding one its string predicate occurrence .
\[
\begin{array}{c}
\pure_{base} {\equiv} \pure \subst{\semp}{\seppred{\code{STR}}{u{,}l{,}r}} {\wedge}l{=}r{\wedge}u{=}\semp  \qquad \fresh~ u_1{,}l_1\\
\pure_{ind} {\equiv} \pure \subst{c{\cdot}\seppred{\code{STR}}{u_1{,}l_1{,}r}}{\seppred{\code{STR}}{u{,}l{,}r}} {\wedge}l_1{=}l{+}1{\wedge}l{<}r {\wedge}u{=}c{\cdot}u_1
\\
\hline
\code{unfold\_str}( \seppred{\code{STR}}{u{,}l{,}r} {\wedge} \pure) ~{\yields}~
 \pure_{base} ~{\vee}~\pure_{ind}
\end{array}
\]
where \form{\pure\subst{t1}{t2}} substitutes all occurrences
of \form{t_2} in \form{\pure} to \form{t_1},
and \form{\fresh ~ \setvars{v}} returns a set of fresh variables
\form{\setvars{v}}.
 The procedure \code{unfold\_str} outputs a set of two disjuncts which are combined from branches of the predicate
\code{STR} with the remainder \form{\pure}.
The constraint \form{u{=}\semp} in base branch
and \form{u{=}c{\cdot}u_1} in inductive branch
are denoted as {\em subterm} constraints. They are
used for constructing a model to witness satisfiability.}

\begin{defn}[Equation Size]
Size of a word equation \form{tr_1{=}tr_2} is the sum of the
notational length of 
 \form{tr_1} and \form{tr_2}.
\end{defn}
We use  \form{\se(n)} to denote a word equation with
size \form{n}.
For example, size of the word equation 
\form{~a~{\cdot}~b~{\cdot}~s{=}s~{\cdot}~b~{\cdot}a~} is 6.

\paragraph{Semantics}
The semantics 
 in this logic is mostly standard.
Every regular expression \form{\regex}
is evaluated to the language \form{\lang(\regex)}.
We define
\[
\begin{array}{lcllcl}
{\SStore}  & {\defsym} &  {\SVar} {\rightarrow} \alphabet^*  & \qquad
{\IStore} & {\defsym} &  {\IVar} ~{\rightarrow}~ Z\
\end{array}
\]
The semantics is given by a forcing relation:
\form{\sstack{,}\istack {\force} \pure}
that forces the interpretation on both string \form{\sstack} and
 arithmetic \form{\istack} to satisfy the constraint
 \form{\pure}
 where \form{\sstack \in {\SStore} }, \form{\istack {\in} {\IStore}},
and \form{\pure} is a formula.

The semantics of our language is formalized as in Figure \ref{prm.sem.fig}.
\begin{figure}[tb]
 \begin{center}
  \begin{minipage}{0.85\textwidth}
  \begin{frameit} \savespace 
\[
\begin{array}{lcl}
\form{\sstack,\istack} {\force} \pure_1 {\vee} \pure_2 & {\iffs} &
 \form{\sstack,\istack} {\force} \pure_1 \text{ or }
 \form{\sstack,\istack} {\force} \pure_2 \\
 \form{\sstack,\istack} {\force} \pure_1 {\wedge} \pure_2 & {\iffs} &
 \form{\sstack,\istack} {\force} \pure_1 \text{ and }
 \form{\sstack,\istack} {\force} \pure_2 \\
\form{\sstack,\istack}  {\force} \neg \pure_1 & {\iffs} &
 \form{\sstack,\istack} \not\force \pure_1 \\
\form{\sstack,\istack} {\force} tr{\in}\regex & {\iffs} & 
\exists w{\in}  \lang(\regex) {\cdot} \sstack,\istack {\force}
 tr{=}w
  \\
\form{\sstack,\istack} {\force} \seppred{\code{STR}}{u{,}n} & {\iffs} & 
\exists w{\in} \alphabet^* {\cdot} \sstack,\istack {\force}
 u{=}w \text{ and }  \istack {\force} |w|{=}n
  \\
 \form{\sstack,\istack} {\force} tr_1{=}tr_2 & {\iffs} & 
 {\sstack}(tr_1){=} {\sstack}(tr_2) \text{ and }
{\istack}(tr_1){=} {\istack}(tr_2)
  \\
\form{\sstack,\istack} {\force} tr_1{\neq}tr_2 & {\iffs} &
\form{\sstack,\istack} {\force} \neg (tr_1{=}tr_2)
  \\
 \form{\sstack,\istack} {\force} {\a_1}{\oslash}{\a_2} & {\iffs} &  {\sstack}(\a_1)~{\oslash}~ {\sstack}(\a_2) \text{, where }
 \oslash \in \{=,\leq\}
  \\
 \end{array}
 \]
 \end{frameit}
 \end{minipage}
\caption{Semantics}\label{prm.sem.fig}
\end{center} 
\end{figure}
We use \form{\true} (\form{\false})
to syntactically denote a valid (unsatisfiable, respectively)
formula.
If \form{\sstack{,}\istack \force \pure},
we use the pair \form{\langle\sstack{,}\istack\rangle} to denote
a solution of the formula \form{\pure}.

 \subsection{Normalized Form}

We would like to remark that word disequalities can be eliminated
using  the approach in
\cite{Abdulla:CAV:2014}. Thus, we only consider
formulas
which contain only one word equation in the normalized form.

 We separate the conjuncts of a formula \form{\pure} into
four parts: \form{\pure{\equiv} \seqs ~{\wedge}~ \astart ~{\wedge}~ \myit{I} ~ {\wedge} ~\subterm } where
(i) \form{\seqs} is a conjunction of  word equations,
(ii) \form{\astart} a conjunction of 
regular expressions,
(iii) \form{\myit{I}} is a conjunction of  arithmetic constraints,
 (iv) and finally \form {\subterm} is a conjunction
 of subterm relations  obtained from
unfolding inductive string predicates.
We notice that if it is unambiguous, we sometimes use \form{\seqs},
\form{\astart}, \myit{I} and \form{\subterm} as sets
instead of conjunctions. And
while string variables
in \form{\seqs} may be encoded with the inductive predicates,
those in \form{\subterm} are not.
For every string inductive predicate \form{\seppred{\code{STR}}{u{,}n}},
its invariant \form{n{\geq}0} must be implied by \form{\myit{I}}.
Each \form{\subterm} is of the form
either \form{s_1{=}c{\cdot}s_2} or \form{s_1{=}s_2{\cdot}s_3}.
They are deduced during solving a formula
and dedicated for constructing a model to witness {\sat}.

\section{Illustrative Example}
 \label{sec.motivate}
 \begin{wrapfigure}{r}{0.3\textwidth} 
\begin{center}
\begin{minipage}{0.28\textwidth} \savespace \saveone
\begin{frameit} \savespace
\begin{tikzpicture}[sibling distance=16mm, level 1/.style={sibling distance=22mm},
     level 2/.style={sibling distance=7mm},
    level distance=22pt,
  every node/.style = {align=center},
   unreach/.style     = {circle,draw,align=center,radius=2mm},
  ]]
  \node {\footnotesize $\D_0^\bigstar$}
      child  { node {\footnotesize \underline{$\D_{11}$} }    }
      child { node {\footnotesize $\D_{12}$}
          child { node {\footnotesize \underline{$\D_{21}$} }}
          child { node {\footnotesize $\D_{22}^\bigstar$} } } 
;
\end{tikzpicture}
\end{frameit}
\end{minipage} 
\caption{Tree $\utree{2}$.}\label{fig.unfolding.tree}
\end{center}\savespace 
\end{wrapfigure}

 We illustrate how {\satsolved} solver
solves satisfiability through the following example:
\[\begin{array}{l}
 \form{\pure{\equiv}a{\cdot}b{\cdot}s{=}s{\cdot}b{\cdot}a~{\wedge}~s{\in}((ab)^*{\cdot}a)~{\wedge}
 |s| {\%} 2{=}0}
\end{array}
\]
Initially, function
\form{\code{init}_{\solvedF}} pairs the string variable \form{s}
 in the word equation with a fresh inductive predicate
 \form{\seppredF{\code{STR}}{u{,}n}} and transforms
the constraint \form{|s|} into a fresh integer variable i.e., \form{n}.
Let
\form{\pure_0~ {=} ~\code{init}_{\solvedF}(\pure)},
 \form{\pure_0} is as follows.
\[
\begin{array}{l}
\pure_0{\equiv} a{\cdot}b{\cdot} \seppredF{\code{STR}}{u{,}n}{=}
\seppredF{\code{STR}}{u{,}n}{\cdot}b{\cdot}a  ~{\wedge}~
 s{\in}((ab)^*{\cdot}a)~{\wedge}~ n\% 2{=}0 ~{\wedge}~ s{=}u
\end{array}
\]
To decide satisifiability, {\satsolved} solver systematically constructs
unfolding trees for the input \form{\pure_0}.
Starting from the unfolding tree $\utree{0}$ with one node \form{\pure_0},
{\satsolved} derives unfolding trees for \form{\pure_0}
 as in Figure \ref{fig.unfolding.tree}.
In this figure, underlined leaves are closed, {\em star} leaves
are linked and $\utree{2}$ is a cyclic proof.
As the word equation in \form{\pure_0} contains inductive predicates,
\form{\pure_0} is not considered for under-approximation.
For over-approximation, {\satsolved} 
replaces every word equations \form{tr_1{=}tr_2}
 by their corresponding length constraints \form{|tr_1|{=}|tr_2|}. As so,
the over-approximation of \form{\pure_0} is:
\form{inv_0{\equiv}2{+}n {=} n{+}2{\wedge}~
 s{\in}((ab)^*{\cdot}a)~{\wedge} n\% 2{=}0 }. Since \form{inv_0}
is not unsatisfiable, {\satsolved} unfolds 
the  predicate instance \form{u} in \form{\pure_0} to
obtain the tree $\utree{1}$
with two leaves \form{\pure_{11}} and \form{\pure_{12}} as follows.
\[
\begin{array}{l}
\pure_{11}{\equiv}\underline{a}{\cdot}b{=} \underline{b}{\cdot}a~{\wedge}~
 s{\in}((ab)^*{\cdot}a)~{\wedge}~n\% 2{=}0 {\wedge} n{=}0 ~{\wedge} s{=}u{\wedge}u{=}\semp\\
\pure_{12}{\equiv}b{\cdot}a{\cdot} \seppredF{\code{STR}}{u{,}n_1}{=}
 \seppredF{\code{STR}}{u{,}n_1}{\cdot}b{\cdot}a~{\wedge}~
 s{\in}((ab)^*{\cdot}a)~{\wedge}
n\% 2{=}0 {\wedge}n{>}0{\wedge} n_1{=}n{-}1~{\wedge} \\
\qquad  \quad s{=}u_1{\wedge}u_1{=}a{\cdot}u
\end{array}
\]
In the $2^{nd}$ iteration,
while
\form{\pure_{11}} is classified as unsatisfiable (unsat cores are underlined),
\form{\pure_{12}} is kept open as \form{\pure_{12}} is not  unsatisfiable.
{\satsolved} unfolds \form{\pure_{12}} to obtain $\utree{2}$ with two leaves as follows.
\[
\begin{array}{l}
\pure_{21}{\equiv}b{\cdot}a{=} b{\cdot}a~{\wedge}~
 s{\in}((ab)^*{\cdot}a)~{\wedge}
\underline{n\% 2{=}0} {\wedge}n{>}0{\wedge} \underline{n_1{=}n{-}1}
{\wedge}\underline{n_1{=}0}
~{\wedge} s{=}u_1{\wedge}u_1{=}a{\cdot}u{\wedge}u{=}\semp\\
\pure_{22}{\equiv}a{\cdot} b{\cdot}\seppredF{\code{STR}}{u{,}l_2{,}r}{=}
 \seppredF{\code{STR}}{u{,}n_2}{\cdot}b{\cdot}a~{\wedge}~
 s{\in}((ab)^*{\cdot}a)~{\wedge}n\% 2{=}0 {\wedge}n{>}0{\wedge} n_1{=}n{-}1{\wedge}\\
\qquad n_1{>}0{\wedge} n_2{=}n_1{-}1~{\wedge} s{=}u_1{\wedge}u_1{=}a{\cdot}u_2{\wedge}u_2{=}b{\cdot}u
\end{array}
\]
In the $3^{rd}$ iteration,
while
\form{\pure_{21}} is marked closed through under-approximation checking,
\form{\pure_{22}} is linked back to \form{\pure_0} by function {\lbsolve}.
{\lbsolve} links \form{\pure_{22}} back to \form{\pure_0} through the following steps.
\begin{enumerate}
\item First, it discards subterm constraints of \form{\pure_0}
and \form{\pure_{22}} as these constraints are for counter-model
construction and not for {\unsat} checking.
Let the remaining formula of \form{\pure_0} and  \form{\pure_{22}} be \form{\pure'_0}
and  \form{\pure'_{22}}, respectively.
\item Secondly, it substitutes the remaining of \form{\pure_{22}} with
the substitution \form{\theta} where \form{\theta=[n'/n,n/n_2]} and
\form{\pure''_{22}{\equiv}a{\cdot} b{\cdot}\seppredF{\code{STR}}{u{,}n}{=}
 \seppredF{\code{STR}}{u{,}n}{\cdot}b{\cdot}a~{\wedge}~
 s{\in}((ab)^*{\cdot}a)~{\wedge}~n'\% 2{=}0 {\wedge}\\
\qquad n'{>}0{\wedge} n'{=}n_1{-}1{\wedge}n_1{>}0{\wedge} n_1{=}n{-}1}.
\item Finally, it checks whether
the string-related part of \form{\pure''_{22}} is identical to its counter-part in
\form{\pure'_0} and arithmetic of \form{\pure''_{22}} implies the arithmetic
of \form{\pure'_0} i.e.,
\[ \form{n'\% 2{=}0 {\wedge}n'{>}0{\wedge}
 n'{=}n_1{-}1{\wedge}n_1{>}0{\wedge} n_1{=}n{-}1 \force n \% 2{=}0}
\]
\end{enumerate}

\section{{\satsolved} Solver}\label{sec.eff.deci}
In this section, we present the semi-decision procedure
{\satsolved}. 
We first describe an overview of {\satsolved}.

\subsection{Overview}\label{sec.generic.sat}

 \begin{algorithm}[t]
  \SetKwInOut{Input}{input}\SetKwInOut{Output}{output}
  \SetAlgoLined

 \Input{$\ind$} \Output{{\sat} or {\unsat}}
   $i {\leftarrow} 0 $; 
  $\ind_0 {\leftarrow} \code{init_{\solvedF}}(\ind)$; 
   $\utree{0} {\leftarrow} \{ \ind_0 \}$ \tcc*{initialize}
 \While { $\true$ }
 {
    (\code{is\_sat},\utree{i}) {$\leftarrow$} \code{\uosolve}(\utree{i})  \tcc*[r]{check {\sat}}
   {\bf if} {\code{is\_sat}} {\bf then }
   {\Return{${\sat}$} \tcc*{\sat}}
   \Else{
     $\utree{i} {\leftarrow} \code{\oosolve}(\utree{i})  $ \tcc*[r]{prune {\unsat}}
      $ \utree{i}  {\leftarrow} \code{\lbsolve}(\utree{i}) $ \tcc*[r]{cyclic proof}
     {\bf if} {\code{is\_closed}(\utree{i})} {\bf then } {\Return{{\unsat} }\tcc*{\unsat}}
	 \Else{
       $\ind_i {\leftarrow} \code{dfs}(\utree{i})  $ ;
	   $ i {\leftarrow} i{+}1 $ \;
       $ \utree{i}  {\leftarrow} {\code{\unfoldsolve}}(\ind_i) $\;
      }
   }
 }
 \caption{{\satsolved} Solver}\label{algo.solved.sat}
\end{algorithm}

The proposed satisfiability solvers {\satsolved} is an
 instantiation of
 the general
satisfiability procedure {\satalgo} presented in
\cite{Loc:CAV:2016}. 
{\satalgo} supports
for a sound and complete base theory (logic) {\basetheory}
augmented with inductive predicates.
The {\em base theory} {\basetheory} must satisfy the following properties: (i)
{\basetheory} is closed under propositional combination and supports
boolean variables; (ii) there exists a complete decision procedure
for {\basetheory}.
We use \form{\base} to denote a formula in {\basetheory}
and \form{\ind} to denote a formula in the extended theory.
Semantically, \form{\ind{\equiv}\bigvee_{i{=}0}^n \base_i, ~ n{\geq}0}.
We remark that in this work the base logic is {\simpleF}
and the extended logic is {\strel} which augmented the base logic
with the inductive predicate \code{STR}.
More inductive predicates to represent recursive functions (i.e., \code{replaceAll})
might be investigated in future work.

The instantiated satisfiability procedure {\satsolved} is presented in Algorithm~\ref{algo.solved.sat}.
Intuitively, to decide satisfiability for a formula, e.g. \form{\pure},
{\satsolved} systematically enumerates
an equivalent set of {\em base} formulas for \form{\pure}.
Particularly, starting from $\utree{0}$ which has one initialized node \form{\ind_0},
 {\satsolved}  iteratively constructs series of {\em unfolding trees} 
$\utree{i}$
for \form{\ind}.
An iteration of the algorithm is described in
lines 3-13.
Function  \code{\uosolve} at line 3 checks whether there exists
a leaf is in base logic and satisfiable.
Function \code{\oosolve} at line 6 over-approximates
a leaf (into the base logic) prior to checking its unsatisfiability.
Function  {\lbsolve} at line 7 links a leaf to an interior node to form
a (partial) cyclic proof.
 Otherwise, it is
marked open.
At line 8, if all leaf nodes are closed,  {\satsolved} returns {\unsat}.
Otherwise,
at line 10 function {\code{dfs}} chooses an open leaf in a
 breadth-first manner and function {\unfoldsolve} unfolds the selected leaf
at line 11.

The construction of cyclic proofs is the most interesting feature of the {\satalgo} framework.
 Intuitively,
a cyclic proof is an
unfolding tree
whose some leaves are marked closed
and remaining leaves are linked back to interior nodes.
Function {\lbsolve} is based on some
weakening and substitution principles \cite{Loc:CAV:2016}.
The soundness of cyclic proof is as follows.

\begin{thm}[\cite{Loc:CAV:2016}]\label{lem.cyclic}
If there is a cyclic proof of \form{\ind}, \form{\ind} is {\unsat}.
\end{thm}

As an instantition of {\satalgo} framework, {\satsolved} is sound for both {\sat} and {\unsat}.
Its soundness is ensured under the following assumptions:
the base logic
{\simpleF} is both sound and complete,
functions \code{\uosolve}, \code{\oosolve} and 
 {\lbsolve}
 are sound,
and function {\code{\unfoldsolve}} has {\em complete} property (i.e.
let \form{\code{\unfoldsolve}(\ind) {\equiv} \ind_1 {\vee} ...{\vee} \ind_k}
then  \form{\ind {\force} \ind_1 {\vee} ...{\vee} \ind_k}).
 {\satsolved} always terminates for {\sat}.
 However, it may, in general, not terminate for {\unsat}.


In the rest of this section, we
define {\simpleF} formulas which is the foundation of the base logic of {\satsolved}.
(subsection \ref{syntax.eff}). Next, in
subsection \ref{solved.algo}
 we present in details
functions of {\satsolved}:  \code{init_{\solvedF}}, \code{\uosolve} (for under-approximation),
 \code{\oosolve} (for over-approximation),
{\lbsolve} (for cyclic proofs) and {\unfoldsolve} (for tree expansion).
 We discuss correctness, termination
and computational complexity results in the next
section.


\subsection{{\simpleF} Fragment}\label{syntax.eff}
 \begin{algorithm}[t]
  \SetKwInOut{Input}{input}\SetKwInOut{Output}{output}
  \SetAlgoLined

 \Input{$(s, \ses)$}\Output{G}
   $G {\leftarrow}  \code{vertex}(s)$;$WL {\leftarrow} \{s\}$\;
 \While { $WL{\not=}\emptyset$ }
 {
    $s_i {\leftarrow}  \code{head}(WL)$;$WL {\leftarrow}  \code{tail}(WL)$\;
    $\code{is\_exist}, tr_i,tr_d,\ses {\leftarrow} \code{choose\_intersect}(s_i,\ses)$\;
    \uIf{\code{is\_exist}}{
      \uIf{$\FV(tr_d)==\emptyset$}{
        \ForEach{$s_j$ $\in$ $\FV(tr_i)$ }{
          $G {\leftarrow}  \code{vertex}(s_j)$\;
          \tcc{mark $s_j$ as leaf}
        }
       }
      \uElse{
         \ForEach{$s_j$ $\in$ $\FV(tr_d)$ }{
          $G {\leftarrow}  \code{vertex}(s_j)$;
          $G {\leftarrow}  \code{edge}(s_i,s_j)$;
          $WL {\leftarrow} WL {\cup} \{ s_j\}$\;
         }
      }
    }
    \uElse{ 
      \tcc{mark $s_i$ as leaf}
    }
 }
 \caption{Dependency Graph Construction}\label{algo.dp}
\end{algorithm}

In this paragraph, we define 
{\simpleF} formulae which are based on
{\em linear formulas} and {\em dependency directed graph}.
\begin{defn}[Linear Formulas]
A formula in {\strel} is said to be linear if it
contains no equality or disequality where a string-typed variable
appears more than once.
\end{defn}

In the following, we present a algorithm
to construct a dependency {\em directed} graph for a conjunction
of word equations.

Let \form{\ses {\equiv} \bigwedge \{ tr_{l_i}{=}tr_{r_i} ~|~ i {\in} 1...n \}} be
 a conjunctive set of word equations.
For each string variable in \form{\ses}, we construct
its dependency graph as in Algorithm \ref{algo.dp}.
This algorithm takes inputs as a pair of variable \form{s}
and a set of equations {\ses}.
It initially generates a graph with one node \form{s}
and a waiting list \form{WL} with one variable  \form{s}.
Function \code{vertex} create a new node if the node does not exist.
In each iteration, it looks for dependent variables of
a variable \form{s_i} in the head of \form{WL}. In particular,
it uses function \code{choose\_intersect} at line 4 to
extract from {\ses} a word equation, e.g. \form{tr_i{=}tr_d},
such that \form{s_i{\in}\FV(tr_i)} (\form{\FV(\pure)} returns free variables
in \form{\pure}). In this case, it returns all variables
in \form{tr_d} as dependent variables of \form{s_i}.
In lines 6-9,
for each word equation
of the form \form{s_1{\cdot}s_2{\cdot}..{\cdot}s_k{=}w} where
 \form{w} is a word in \form{\alphabet^*},
we mark \form{s_1}, \form{s_2}, .., \form{s_k}
as leaves.
We remark that when a node is marked as leaf,
its out-going edges are removed and it is never added
into the waiting list.
Otherwise, it adds a directed edge
from \form{s_i} to a dependent node \form{s_j} using function \code{edge}.
We notice that there may be more than one edge between two nodes.

\begin{defn}[{\simpleF} Formulas]
A formula \form{\pure} 
is said to be in
{\simpleF} fragment if \form{\pure} is linear
and for all dependency graphs \form{G} built for each string variable
in \form{\pure}, \form{G} does not contain any cycle.
\end{defn}

We find that {\simpleF} fragment is equivalent
to the acyclic form presented in 
\cite{Abdulla:CAV:2014}, and thus
satisfiability problem for {\simpleF} formulas
is decidable.  We explicitly state
this decidability as follows.
\begin{thm}[{\simpleF} Decidability \cite{Abdulla:CAV:2014}]
The satisfiability problem for {\simpleF} is decidable.
\end{thm}

\subsection{{\satsolved} Instantiation}\label{solved.algo}


The satisfiability procedure {\satsolved} is an instantiation 
of the generic framework {\satalgo} presented in Algorithm~\ref{algo.solved.sat}.
{\satsolved} takes a formula \form{\ind} as input,
initially pairs each {\em bare} string variable
in word equations with a fresh string {\em inductive} predicate.
(using function \code{init_{\solvedF}}), and then
systematically enumerates disjuncts \form{\base_i}.
{\satsolved} can produce
two possible outcomes: {\sat} with
a model obtained from a satisfiable 
formula
\form{\base_i}
 or {\unsat} with a proof; 
non-termination is classified as
 \code{UNKNOWN}.
We recap that while our discussion focuses on formulas with only
string equalities, a string disequality can be reduced
to a finite set of equalities. An implementation for such reduction can be found
in \cite{Abdulla:CAV:2014}.

In the rest of this subsection,
we present the base logic and instantiation of
functions \code{init_{\solvedF}}, \code{\uosolve},
 \code{\oosolve},
{\lbsolve}, and {\unfoldsolve}.

\paragraph{Base Logic}
The base fomulae of {\satsolved} is defined as follows.

\begin{defn}[Base Formula] Let \form{\pure{\equiv} \seqs {\wedge} \astart {\wedge} \myit{I}  {\wedge} \subterm}.
\form{\pure} is a base formula of solver {\satsolved}
if it is in fragment {\simpleF}
and \form{\seqs} 
does not contain any inductive predicate instance.
\end{defn}
We use \form{\satb{\base}} to denote the satisfiability checking
for base formula \form{\base}.
Both function {\uosolve} and {\oosolve}
 invoke \form{\satb{...}} to discharge base formulas.

\paragraph{Initializing}
Let \form{\pure{\equiv}\seqs {\wedge} \astart {\wedge} \myit{I} {\wedge} \subterm} be the input.
Function \code{init_{\solvedF}} pairs each string variable in \form{\seqs} 
 with a
 predicate instance \code{STR}.
In particular, for each variable
\form{s_i}, 
we generate a new inductive predicate
 \form{\seppred{\code{STR}}{u_i{,}n_i}} where
\form{u_i} and \form{n_i} are fresh variables,
 conjoins the constraint \form{s_i{=}u_i} into \form{\subterm}, and
conjoins  a conjunction of
invariant of each length function \form{\bigwedge \{ n_i{ \geq}0 \}}
into \form{\myit{I}}.
After that, we replace all length function of \form{s_i},
i.e., exhaustively reduce all expression \form{|tr_i|} and then substitue
each \form{|s_i|} expression
in \form{\myit{I}} by the corresponding variable \form{n_i}.

\paragraph{Approximating}
For soundness of {\sat}, under-approximation function {\uosolve} only
 considers {\em base} leaves, those leaves which are in the base logic. 
Over-approximation function {\oosolve} reduces each leaf with inductive predicates to a base formula
 by replacing each word equation \form{tr_1{=}tr_2} with the corresponding length constraint
 \form{|tr_1|{=}|tr_2|}.
For example, the following formula
\[ \form{\pure~{\equiv}~\seppredF{\code{STR}}{u{,}n_u}{=}
\seppredF{\code{STR}}{v{,}n_v}{\cdot}\seppredF{\code{STR}}{u{,}n_u}{\cdot}a{\cdot}\seppredF{\code{STR}}{u{,}n_u}{\cdot}\seppredF{\code{STR}}{t{,}n_t} ~{\wedge}~n_u{\geq}0{\wedge}n_v{\geq}0{\wedge}n_t{\geq}0}
\]
is over approxiamted into 
\form{\overline{\pure}~{\equiv}~n_u{=}n_v{+}n_u{+}1{+}n_u{+}n_t ~{\wedge}~n_u{\geq}0{\wedge}n_v{\geq}0{\wedge}n_t{\geq}0}.
\form{\overline{\pure}} is passed to \form{\satb{...}} to check its satisfiability.
As \form{\overline{\pure}} is unsatifiable, so is \form{{\pure}}.

\paragraph{Expanding}
{\satsolved} chooses an open leaf, e.g. node \form{i}, in
a depth-first manner
(at line 10 of Algorithm \ref{algo.solved.sat}) and unfolds it using function {\unfoldsolve}
The function {\unfoldsolve} chooses one word equation of the node \form{i},
e.g. \form{tr_{l_i}{=}tr_{r_i}},
and examines two {\em leading} terms
at the head of \form{tr_{l_i}} and \form{tr_{r_i}}.
After that, it unfolds a predicate instance \code{STR}
 accordingly, matches/consumes
and returns a set \form{L} of formulas.
If this set is empty, the algorithm marks the node \form{i} closed.
Otherwise, for each formula in \form{L} it creates a new node \form{j} and
new edge from \form{i} to \form{j}.


Function {\unfoldsolve} is the core of our algorithm.
It aims to reduce word equations
to base disjuncts.
 Intuitively, it
applies Unfold-and-Match
on the leading (first) term (string variable or character constant) of both sides of an equation.
In particular, this function examines the following
three cases. 

\noindent{\bf Case 1.}
In this case,
 the leading terms at LHS and RHS
are characters in the alphabet. It then {\em matches} these two
characters, reduces the size of the word equation and
 thus makes progressing.  Two subcases are formalized as follows.
\[
\begin{array}{c}
\unf{\solvedF-CONST-SUCC}\\
\code{unfold}(tr_1{=}tr_2 {\wedge} \pure) ~{\yields}~ L
\\
\hline
\code{unfold}( c{\cdot}tr_1{=}c{\cdot}tr_2 {\wedge} \pure) ~{\yields}~L
\end{array}
\quad \quad
\begin{array}{c}
\unf{\solvedF-CONST-FAIL}\\
c_1{\not=} c_2
\\
\hline
\code{unfold}( c_1{\cdot}tr_1{=}c_2{\cdot}tr_2 {\wedge} \pure) ~{\yields}~\{~\}
\end{array}
\]
In  the first sub-case (rule  $\unf{\solvedF-CONST-SUCC}$),
 these two terms are identical;
function {\unfoldsolve} consumes them and makes progressing.
In  the second sub-case (rule $\unf{\solvedF-CONST-FAIL}$),
 these two terms are not identical;
function {\unfoldsolve} returns an empty set
and classifies this leaf
unsatisfiable.

\noindent{\bf Case 2.} In the second case, 
 one
 leading term is a character \form{c}
and another is a  predicate instance \form{\seppred{\code{STR}}{u{,}n}}.
This case is formalized by the folowing two rules
corresponding two cases where the inductive predicate
is in LHS ($\unf{\solvedF-SMALL-L}$)
or RHS ($\unf{\solvedF-SMALL-R}$).
\[
\begin{array}{c}
\begin{array}{c}
\unf{\solvedF-SMALL-L}\\
\seqs_1 {=}  (tr_1{=}c{\cdot}tr_2{\wedge}\seqs)\subst{\semp}{\seppred{\code{STR}}{u{,}n}}\quad
\myit{I}_1{=}\myit{I} {\wedge}n{=}0 \quad
\subterm_1{=}\subterm {\wedge} u{=}\semp \\
\seqs_2 {=} (\seppred{\code{STR}}{u{,}n_1} {\cdot} tr_1{=}tr_2{\wedge}\seqs) \subst{c{\cdot}\seppred{\code{STR}}{u{,}n_1}}{\seppred{\code{STR}}{u{,}n}} \\
\fresh~u_1,n_1 \quad
 \myit{I}_2{=} \myit{I} {\wedge}n_1{=}n{-}1{\wedge}n{>}0
\quad  \subterm_2{=} \subterm \subst{u_1}{u}{\wedge} u_1{=}c{\cdot}u \\
\hline
{\unfoldsolve}(\seppred{\code{STR}}{u{,}n}{\cdot}tr_1{=} c{\cdot}tr_2 {\wedge} \seqs {\wedge} \astart   {\wedge}  \myit{I} {\wedge} \subterm)
{\yields}
\{ \seqs_1 {\wedge} \astart   {\wedge}  \myit{I}_1 {\wedge} \subterm_1;
  \seqs_2 {\wedge}  \astart   {\wedge}  \myit{I}_2 {\wedge}\subterm_2\} \\
\end{array} \\
\begin{array}{c}
\unf{\solvedF-SMALL-R}\\
\seqs_1 {=} (c{\cdot}tr_1{=}tr_2{\wedge}\seqs) \subst{\semp}{\seppred{\code{STR}}{u{,}n}}\quad
\myit{I}_1{=}\myit{I} {\wedge}n{=}0 \quad
\subterm_1{=}\subterm {\wedge} u{=}\semp  \\
\fresh~u_1,n_1 \quad
\seqs_2 {=} (tr_1{=} \seppred{\code{STR}}{u{,}n}{\cdot}tr_2{\wedge}\seqs) \subst{c{\cdot}\seppred{\code{STR}}{u{,}n_1}}{\seppred{\code{STR}}{u{,}n}} \\
 \myit{I}_2{=} \myit{I} {\wedge}n_1{=}n{-}1{\wedge}n{>}0
\quad  \subterm_2{=} \subterm \subst{u_1}{u}{\wedge} u_1{=}c{\cdot}u \\
\hline
{\unfoldsolve}(c{\cdot} tr_1{=} \seppred{\code{STR}}{u{,}n}{\cdot}tr_2 {\wedge}\seqs {\wedge}  \astart   {\wedge}  \myit{I} {\wedge} \subterm)  {\yields}
\{\seqs_1 {\wedge} \astart   {\wedge}  \myit{I}_1 {\wedge} \subterm_1 ;
\seqs_2 {\wedge}  \astart   {\wedge}  \myit{I}_2 {\wedge}\subterm_2\} \\
\end{array}
\end{array}
\]
In these rules,
function {\unfoldsolve} does case split by unfolding the predicate to
consider two cases: \form{u} is an empty word or it is a word whose
the first character is \form{c}.
In the latter case, our system substitutes \form{\seppred{\code{STR}}{u{,}n}}
by the concatenation \form{c \cdot \form{\seppred{\code{STR}}{u{,}n_1}}}
where \form{n_1{=}n{-}1}. The reuse of variable \form{u}
is critical to identify back-links in the unfolding trees.
After this {\em selectively} {\em unfolding}, {\unfoldsolve}
matches the character in both sides and makes progressing (i.e., reducing the size of the word equations).

\noindent{\bf Case 3.} In the last case,  the leading terms on both LHS and RHS are
 inductive predicate instances, e.g. \form{\seppred{\code{STR}}{s_1{,}n_1}}
and \seppred{\code{STR}}{s_2{,}n_2}.
\[
\begin{array}{c}
\renewcommand{\arraystretch}{1.1}
\unf{\solvedF-BIG}\\
\seqs_1 {=} (tr_1{=}tr_2 {\wedge}\seqs) \subst{u_1}{u_2}\quad
\myit{I}_1{=}\myit{I} {\wedge}n_1{=}n_2 \quad
\subterm_1{=}\subterm {\wedge} u_1{=}u_2 \quad
 L_1{=}\{\seqs_1 {\wedge} \astart   {\wedge}  \myit{I}_1 {\wedge} \subterm_1\} \\
\seqs_2 {=} (\seppred{\code{STR}}{u_1{,}n_3}{\cdot}tr_1{=}tr_2{\wedge}\seqs) \subst{\seppred{\code{STR}}{u_2{,}n_2} {\cdot}\seppred{\code{STR}}{u_1{,}n_3}}{\seppred{\code{STR}}{u_1{,}n_1}} \\
\myit{I}_2{=}\myit{I} {\wedge} n_3{=}n_1{-}n_2 \quad
 \subterm_2{=}\subterm \subst{u_3}{u_1}{\wedge} u_3{=}u_2 {\cdot}u_1 \quad
L_2{=}\{ \seqs_2 {\wedge} \astart   {\wedge}  \myit{I}_2 {\wedge} \subterm_2\}\\
\seqs_3 {=} (tr_1{=}\seppred{\code{STR}}{u_2{,}n_3}{\cdot}tr_2{\wedge}\seqs) \subst{\seppred{\code{STR}}{u_1{,}n_1} {\cdot}\seppred{\code{STR}}{u_2{,}n_3}}{\seppred{\code{STR}}{u_2{,}n_2}} \\
\myit{I}_3{=}\myit{I} {\wedge} n_3{=}n_2{-}n_1 \quad
 \subterm_3{=}\subterm \subst{u_3}{u_2}{\wedge} u_3{=}u_1 {\cdot}u_2 \quad
 L_3{=}\{\seqs_3 {\wedge} \astart   {\wedge}  \myit{I}_3 {\wedge} \subterm_3\} \\
\hline
{\unfoldsolve}(\seppred{\code{STR}}{u_1{,}n_1}{\cdot}tr_1{=} \seppred{\code{STR}}{u_2{,}n_2}{\cdot}tr_2
{\wedge}\seqs {\wedge}  \astart   {\wedge}  \myit{I} {\wedge} \subterm) ~{\yields}~
L_1 {\cup} \{ L_2\} {\cup} \{ L_3\}
\end{array}
\]
Function {\unfoldsolve} expands the tree
through a {\em big-step} unfolding.
As shown in rule $\unf{\solvedF-BIG}$,
it considers the following three subcases: (i) two string variables are identical
(i.e., \form{u_1{=}u_2} in the first line); (ii) \form{u_2} is a substring of
\form{u_1} (i.e., \form{u_1} is substitued by \form{u_2 {\cdot}u_1} in the second and third lines); and (iii)
\form{u_1} is a substring of
\form{u_2} (i.e.,  \form{u_2} is substituted by \form{u_1 {\cdot}u_2} in the fourth and fifth lines).
We notice that while the first subscase make progressing (i.e., reducing the size of the word equations),
 the remaining two cases
do not.

\paragraph{Linking Back}
Function {\lbsolve} attempts to link remaining open leaves back
to interior nodes so as to form
 a fixpoint (i.e., a pre-proof for induction proving) \cite{Loc:CAV:2016}.
This function is implemented through some weakening
and substitution principles.
In particular, function {\lbsolve}
links a leaf to an interior node if after some substitution,
 (i) the leaf has
isomorphic word equations and regular membership to the inter node;
and (ii) its arithmetical part implies
the arithmetical part of the inter node.
We notice that the subterm constraints in each leaf are for 
counter-model construction
and are discarded during this linking.
The substitutions are identified based on
isomorphic string terms
 and {\em well-founded} ordering relations \form{R} over arithmetical
variables.
In the following, 
we define isomorphic relation between word equations.
The isomorphic relation between regular expression
is similar.
\begin{defn}[isomorphic equations]
The equations \form{\se_1} and \form{\se_2} are isomorphic
if \form{\se_1} and \form{\se_2} become identical when we replace
all string variables \form{u} in \form{\se_1} by \form{\code{permute}(u)}
and all characters \form{c} in \form{\se_1} by \form{\code{permute}(c)},
where \form{\code{permute}(u)} is a permutation function on \form{\SVar},
and \form{\code{permute}(c)} is a permutation function on the
alphabet \form{\alphabet}.
\end{defn}

In the next section, we will describe a decidable subfragment
which includes arithmetic based on
classes of well-founded ordering relations.

\section{Correctness}\label{solved.correct}
In this section, we discuss the soundness and termination of our solver.
 We also
provide a complexity analysis of our decision procedure
to show that it
runs, in the worst case, in linear time for {\simpleF}
and factorial time for {\solvedF}.
\subsection{Soundness}
The soundness of our {\satsolved} algorithm
relies on the correctness of functions {\uosolve}, {\oosolve}
and {\unfoldsolve}. 
The soundness of functions {\uosolve} and {\oosolve} is straightforward.
Additionally, it is easy to verify that our unfolding rules
have 
the {\em complete} property.
We state the correctness of the proposed {\satsolved} algorithm as follows.

\begin{thm}[Soundness]
Let $\utree{i{+}1}$ be the unfolding tree obtained after expanding the tree 
$\utree{i}$ using function \code{\unfoldsolve}. Then
\begin{itemize}
\item $\utree{i}$ has a {\sat} leaf with a solution \form{\langle\sstack{,}\istack\rangle} implies that there exists 
\form{\sstack{\subseteq}\sstack'} and
\form{\istack{\subseteq}\istack'} such that $\utree{i{+}1}$ has a {\sat} leaf with solution \form{\langle\sstack'{,}\istack'\rangle}.
\item $\utree{i{+}1}$ has a {\sat} leaf with
a solution \form{\langle\sstack'{,}\istack'\rangle} implies that
 $\utree{i}$ has a {\sat} leaf  with a solution \form{\langle\sstack{,}\istack\rangle} where \form{\sstack{\subseteq}\sstack'} and
\form{\istack{\subseteq}\istack'}.
\end{itemize}
\end{thm}

\subsection{Decidable Fragment}
In this section, we show that our solver terminates
for the subfragment {\solvedF} which is defined as follows.

\paragraph{{\solvedF} Formulae}
The arithmetical constraints over length functions of
{\solvedF} formulas are restricted
on periodic relations \form{R} \cite{Marius:CAV:2010} which is defined
as follows.
For each string variable \seppredF{\code{STR}}{u_i{,}x_i}, 
 let \form{u'_i} (\seppredF{\code{STR}}{u'_i{,}x'_i}) be subterm of \form{u_i} where \form{x'_i{>}x_i}. 
Finally, let \form{\setvars{x}} and \form{\setvars{x}'}
be sequences of \form{k} such \form{x_i} variables.
\form{R \in Z^k \times Z^k} is an integer relation over
variables \form{\setvars{x}} and \form{\setvars{x}'}, its
{\em transitive closure} \form{R^+{=}\bigcup_{i{=}1}^\infty}
where \form{R^1{=}R} and \form{R^{i{+}1}{=}R^i{\circ}R}
for all \form{i{\geq}1}.
Relation \form{R} is defined as one of the two following form.
\begin{itemize}
\item Octagonal relation. An {\em octagonal relation} is a finite
conjunction of constraints of the form
 \form{R(x_1{,}x_2){\equiv}\underline{+} x_1 \underline{+} x_2{\leq}k}
where \form{k} is an integer constraint, 
\form{x_1{,}x_2 {\in} \setvars{x} {\cup}\setvars{x}'}.
\item Finite linear affine relation. A {\em linear affine relation}
is a finite
conjunction of constraints of the form
 \form{R(\setvars{x}{,}\setvars{x}'){\equiv}C\setvars{x} {\geq}D \wedge \setvars{x}'{=}A\setvars{x}+B}, where \form{A\in  Z^{k\times k}},
\form{C\in  Z^{p\times k}}  are matrices,
and \form{B\in  Z^k},  \form{D\in  Z^p}.
A linear affine relation is finite if the set
\form{\{ A^i \mid  i{\geq}0 \}} is finite.
\end{itemize}
For example \form{R(x_1{,}x_2){\equiv}x_1-x_2{=}5} is an octagonal relation
as it is equivalent to\\
 \form{R(x_1{,}x_2){\equiv}x_1-x_2{\leq}5 \vee x_2-x_1{\leq}-5}. Especially,
the authors in \cite{Marius:CAV:2010} show that
the transitive closure of these periodic relations is Presburber-definable
 and effectively computable. In other words, these relations
are ultimately periodic. The set of periodic is defined as follows.
\begin{defn}
A set $S$ of integers is defined to be {\em ultimately periodic} if
there are some $M \ge 0$, $p > 0$ such that
$n \in S$ iff $n+p \in S$ for all $n \ge M$.
Then we call the set $(M,p)$-periodic.
\end{defn}
The set $(M,p)$-periodic is important for the complexity analysis.

\begin{defn}[{\solvedF} Formulae]
A formula \form{\pure} 
 is said to be in
{\solvedF} fragment if either
it is in {\simpleF} subfragment
or it satisfies the two following restrictions
\begin{enumerate}
\item For all dependency graph \form{G} built for each string variable
in \form{\pure}, \form{G} contains at most one cycle, including self-cycle.
\item  zero or more arithmetical periodic constraints  \cite{Marius:CAV:2010}
(as defined above)
 on the length functions 
of string variables. 
\end{enumerate}
\end{defn}


\paragraph{Termination and Complexity}
Function {\unfoldsolve} produces
a set of new formulas whose either
i) size are decreased
or
ii) all
variables in the chosen word equation are suffix
of the corresponding in the input
and there is at least one variable is strict suffix.
Hence, {\satsolved} procedure always terminates for {\sat}.
The substitution in the rules
\form{\unf{\solvedF-SMALL-*}} and
\form{\unf{\solvedF-BIG}} may infinitely increase the
sizes 
of word equations when these equations include multiple
occurrences of one variable. Thus,
in general, {\satsolved} algorithm
may not terminate. In the following,
we show that {\satsolved} indeed terminates
for {\simpleF} and {\solvedF} formulas.
We also provide computational complexity analyses.

\begin{thm}\label{thm.simpleF}
Let \form{\pure{\equiv} \seq_1 {\wedge}...{\wedge}\seq_M {\wedge} \astart   {\wedge} \myit{I} {\wedge} \subterm} be in the {\simpleF} fragment.
{\satsolved} terminates for \form{\pure}.
If \form{M} word equations are of the form \form{tr_{l_i}{=}tr_{r_i}} where \form{i \in \{1..M\}},
and \form{N} is the longest notational length
of these word equations,
 then
the length of every path of the derived unfolding trees for \form{\pure}
is $\mathcal{O}(\form{2^MN})$.
\end{thm}
\begin{proof}
As \form{\pure} is in {\simpleF},
it is linear as well as there no cycle in dependency graphs
derived for its every string variables.
As \form{\pure} is linear, the size of the
word equation obtained from unfolding the word equation
{\se} is less than the size of {\se}.
Furthermore, as there is no cycle in any dependency graph,
the formulas after the substitution while
 unfolding using either rule
 \form{\unf{\solvedF-SMALL-*}} or rule \form{\unf{\solvedF-BIG}},
are still linear.
Thus, \form{\pure} is reduced to a set of base formulas
in finite steps.

We remark that after each unfolding on the word equation
{\se}, while the size of result decreases at least one,
the size of the each remaining word equation
in {\ses} increases at most one.
Thus, whenever reducing one word equation
to size 0,  size of
 each remaining word equations in {\ses} increases
$\mathcal{O}(N)$. Based on this fact, the complexity is
$\mathcal{O}(\form{(1{+}2^1{+}2^2+...+2^{(M-1)})N})$.
Indeed, we can prove the computational complexity
above by induction on \form{M}.
\qed
\end{proof}

This theorem implies that {\satsolved} solves
a {\simpleF} with one word equation, in the worst case, in linear time.
In the next theorem, we show that {\solvedF} indeed terminates
for a formula with multiple word equation.
\begin{thm}[{\solvedF} Termination]
Let \form{\pure{\equiv}\seq_1{\wedge}..{\wedge}\seq_M {\wedge} \astart {\wedge} \myit{I} {\wedge}\subterm} 
 be in the {\solvedF} fragment.
{\satsolved} terminates for \form{\pure}.
\end{thm}
\begin{proof}
The proof for the formula in {\simpleF} is given in Theorem \ref{thm.simpleF}.
In the following, we consider the formula which is in another case.
We remark that
unfolding rules of function \code{unfold}
decrease the size of on-processing (the first one in these rules)
word
equation at least one and increases
 the size of each remaining equation in \form{\seqs}  at most
one during the substitution.
As the input formula \form{\pure} is in the {\solvedF} fragment,
neither
(i) this on-processing equation
 includes any string variable which occurs
more than twice
nor (ii) any dependency graphs derived for variables
of \form{\pure} contains more than one loop.
(i) guarantees that size
of the on-processing word equation after unfolded
is never longer than the size of original equation.
(ii) ensures that \form{\pure}
is still in  the {\solvedF} fragment after
the substitution.
As a permutation of a word equation with a given
length is finite, these equations
are isomorphic to an inner node
after a finite number of unfoldings.
We notice that, in these rules
$\unf{\solvedF-SMALL-*}$ and $\unf{\solvedF-BIG}$,
the new subterm constraints are generated on length functions and
they are \form{R} {\em periodic} relations which are Presburger definable.
This means they can be reduced to an equivalent Presburger constraints
in finite time. Hence, function {\lbsolve} can always link back
every leaves after a finite number of unfoldings.
Thus, {\satsolved} terminates for a {\solvedF} formula.
\qed
\end{proof}

Finally, we state the computational complexity of the satisfiability problem
for {\solvedF}. For simplicity, we only discuss
 the case where \form{\pure} contains one word equation.
The proof for the complexity relies on the following lemma
which states that given a periodic relation corresponding a set \form{S},
any formula derived from the unfolding of this relation corresponds to
a set \form{S'} and \form{S' \subseteq S}.

\begin{lemma}\label{lemma:periodic}
If $S \ne \emptyset$ is $(M,p)$-periodic
and $S'{=}\{ y \ |\  y = kpx{,}~ x \in S \}$, then $S'$ is $(M,kp)$-periodic and \form{S' \subseteq S} for k is an integer and $k >0$.
\end{lemma}
It is easy to show that if \form{x \in S} and \form{x{\ge}M}, then
 \form{kpx{\ge}M}, \form{kpx \in S'} and \form{kpx+kp \in S'}.

\begin{thm}[{\solvedF} Complexity]
Let \form{\pure{\equiv}tr_l{=}tr_r {\wedge} \astart {\wedge} \myit{I} {\wedge}\subterm} 
 be in the {\solvedF} fragment.
The length of every path in the derived unfolding trees for \form{\pure}
is $\mathcal{O}(N^2(N!))$ where N is the size 
of the equation \form{tr_l{=}tr_r}.
\end{thm}
\begin{proof}
This complexity result is based on the following four facts.
\begin{enumerate}
 \item
 Size of a word equation of any node
 in the derived unfolding trees for \form{\pure}
is less than or equal \form{N};
Hence, there are $\mathcal{O}(N)$ possibilities for the length.
\item
 There are $\mathcal{O}(N!)$ possibilities to arrange a sequence of \form{N} symbols of the respective either string variables or characters.
 \item 
 For every arrangement, i.e. a word equation, there  are $\mathcal{O}(N)$ possibilities to distinguish two sides (LHS and RHS) of the equation.
 \item
In a path, arithmetical part of a {\satsolved} formula is a disjunct of
the unfolding from its descendant.
From lemma \ref{lemma:periodic}, the set \form{S'} of this disjunct
is a subset of set corresponding its descendant.
Thus, the function {\lbsolve} can always link the arithmetical part of
such above leaf to any its descendant nodes.
\end{enumerate}
\qed
\end{proof}



\section{Related Work} \label{sec.related}

Makanin notably provided a mathematical proof for the satisfiability problem
of word equation \cite{Makanin:math:1977}.
In the sequence of papers, Plandowski {\em et.al.} showed that the complexity of
this problem is PSPACE 
\cite{Jaffar:JACM:1990,Plandowski:FOCS:1998,Plandowski:FOCS:1999,Plandowski:STOC:2006}.
Beside the development of the foundation for the {\em acyclic} form
\cite{Abdulla:CAV:2014,Liang:FroCos:2015}
as discussed in section \ref{sec.intro},
Ganesh {\em et. al.} presented undecidability result
for quantified string-based formulas \cite{Ganesh:HVF:2012}.
In the rest of this section,
we summarize the development of related works on practical string solvers.

\paragraph{Automata-based Solvers.} Finite automata provides a natural encoding
for string with regular membership constraints.
 Rex \cite{Veanes:ICST:2010} encodes strings as symbolic finite automata (SFA).
Each SFA transition is transformed into SMT constraints.
Java String Analyzer (JSA) \cite{Christensen:SAS:2003} is specialized
for Java string constraints. JSA approximates string constraints 
into multi-level automaton.
 \cite{Hooimeijer:PLDI:2009,Hooimeijer:ASE:2010}
provides a reasoning over string with {\em priori} length bounds.
Recent work in \cite{Aydin:CAV:2015} provides
a length-bound approach for solving string constraints and further
counting the number of solution to such constraints.
Recently, authors in \cite{Abdulla:CAV:2014,Abdulla:CAV:2015}
proposes a DPLL(T)-based approach to
unbounded string constraints with regular expressions and length function.
 \cite{Wang2016} described
a new method based on a scalable logic circuit representation to support
 various string and automata manipulation operations and
 counter-example generatation.
In our view, inductive predicate could represent automaton. Thus,
tt is interesting
to investigate how we could
adapt the proposed algorithm {\satsolved}
for the prolems based on automata.

\paragraph{Bit-vector-based Solvers.}
Hampi solver \cite{Kiezun:ISSTA:2009}
reduces fixed-sized string constraints to bit-vector problem
and then satisfiability.
The Kazula solver \cite{Saxena:SP:2010} extends Hampi with
{\em concatenation} operation. It
first solves arithmetical constraints and then
 enumerates possible fixed-length versions
of an input formula using Hampi.
In \cite{Bjorner:TACAS:2009}, strings are represented as arrays.
Discharging string with length constraints are performed through two phases.
First an integer-based over-approximation of the string constraint is solved and then 
fixed-length string constraints are then decided in a second phase.

\paragraph{Word-based Solvers.}
Z3str \cite{Zheng:FSE:2013} implements string theory as an extension
of Z3 SMT solver through string plug-in. It supports unbounded string constraints with
a wide range of string operations. Intuitively,
it solves string constraints and generates string lemmas
to control with z3's congruence closure core.
Z3str2 \cite{Zheng:CAV:2015} improves Z3str by
proposing a detection of those constraints beyond the tractable fragment,
 i.e.  overlapping arrangement,
and pruning the search space for efficiency.
Similar to Z3str, CVC4-based string solver \cite{Liang:CAV:2014}
communicates with CVC4's equality solver to exchange information over string.
S3 \cite{Trinh:CCS:2014} enhances Z3str
to incrementally interchange information between string and arithmetic constraints.
S3P \cite{Trinh:CAV:2016} further extends
S3 to detect and prune non-minimal subproblems
while searching for a proof.
While the technique in S3P aims
for satisfiable formulae, it may returns unknown for unsatisfiable formulas
due to absence of multiple occurrences of each string variable.
Our solver can support well for both classes of queries in case of less than
or equal to two  occurrences of each string variable.

\section{Conclusion and Future Work}
We have presented the semi-decision procedures
 {\satsolved} for the problem of solving
satisfiability of a {\strel} formula with
word equations, regular membership
and length functions.
We have shown that the proposed procedure terminates
for the subfragment {\simpleF}
and provided its computational complexity.


For future work, we would like to implement
the proposed decision procedure {\satsolved} based on the generic
{\satalgo} framework \cite{Loc:CAV:2016}.
As the {\satalgo} framework naturally supports arbitrary user-defined
predicates,
we might extend the proposed decision procedure with
inductive predicates encoding 
 recursive string functions  (i.e., function \code{replace}) \cite{Trinh:CAV:2016}.
\hide{ For example,
Kleene star operation
can be represented as inductive predicate as follows.
For example, \form{s\in (ab)^*} is encoded as
\form{s{=}u {\wedge} \seppred{\code{STAR}}{u{,}``ab"{,}l{,}r}}
where the inductive predicate \form{\seppred{\code{STAR}}{u{,}w{,}l{,}r}}
is defined as follow.
\[
\begin{array}{l}
\defpred~ \seppred{\code{STAR}}{u{,}w{,}l{,}r}  ~{\equiv}~ u{=}\semp {\wedge}l{=}r \\
\qquad~{\vee}~ ~ \seppred{\code{STR}}{u_1{,}w{,}l_1{,}r}
 {\wedge}u{=}w{\cdot}u_1 {\wedge}
 l_1{=}l{+}|w| {\wedge} l{<}r. \\
\end{array}
\]}
We were hoping that such extension helps enhance the completeness 
of the string logic augmented with these recursive functions.



\bibliographystyle{abbrv}
\bibliography{all}

\rep{
\newpage
 \appendix
}
\end{document}